\documentclass[final,5p,times,twocolumn]{elsarticle}
\usepackage{graphicx}
\usepackage{slashed,color}
\usepackage{amsmath,amsfonts,bm}
\usepackage{amssymb}
\usepackage{calligra}
\usepackage[T1]{fontenc}

\newcommand{\nn}{\nonumber \\ }

\journal{Physics Letters  B}

\begin{document}



\author{A.~V.~Radyushkin}
\address{Physics Department, Old Dominion University, Norfolk,
             VA 23529, USA}
\address{Thomas Jefferson National Accelerator Facility,
              Newport News, VA 23606, USA
}

\title{Target  Mass Effects in   Parton Quasi-Distributions }

\begin{abstract}

We study the impact  of   non-zero (and apparently  
large)  value of the nucleon mass $M$ on the shape of  parton quasi-distributions
 $Q(y,p_3)$, in particular on its  change with  the change of the 
nucleon momentum 
$p_3$. We observe  that the usual  target-mass corrections 
induced by the $M$-dependence of the twist-2 operators are 
rather small. Moreover, we show that within  the framework
based on parametrizations by  transverse momentum dependent 
distribution functions (TMDs) these corrections are 
canceled by  higher-twist contributions.
We identify a novel  source of kinematic target-mass dependence 
of TMDs and build models corrected for such  dependence.
We find that  resulting changes   may be safely neglected 
for $p_3 \gtrsim 2M$.

\end{abstract}

\maketitle


\section{Introduction}

The parton quasi-distributions (PQDs) $Q(y,p_3)$
recently  proposed by X. Ji  \cite{Ji:2013dva} convert into  usual twist-2 parton distribution 
functions (PDFs) $f(y)$ when the hadron momentum $p_3$ tends to infinity.
Unlike PDFs that are  defined through a correlator of quark fields separated by 
a light-like interval $z$, the definition of QPDs  refers to  the interval $z$ that
has only a space $z_3$ component. This opens a possibility 
to extract PQDs from Euclidean lattice gauge calculations. 

It is expected that,  for a finite $p_3\equiv P$, the difference between $Q(y,P)$ and $f(y)$ 
is  explained  by the higher-twist  and target-mass corrections in powers of $\Lambda^2/P^2$  
and $M^2/P^2$, respectively. 

The target-mass dependence of the twist-2 matrix elements is well-known since mid 70's
\cite{Nachtmann:1973mr,Georgi:1976ve}. In Ref. \cite{Lin:2014zya}    
(see also \cite{Alexandrou:2015rja,Monahan:2016bvm,Zhang:2017bzy}), 
 this information   was used 
to connect   $x^n$ moments of PDFs $f(x)$ and  $y^n$ moments of  the  twist-2 part of QPDs $Q(y,P)$.  In Ref.  \cite{Chen:2016utp}, this  connection was converted into   
 a direct relation
between the functional forms of  $Q_{\rm twist-2} (y,P)$ and  $f(y)$.  
In the present paper, we give our derivation 
of this relation and emphasize that 
for $y>0$ and $y<0$ components of  $f(y)$, it  reduces to    a simple rescaling
by factors depending on the ratio $M^2/P^2$.  

We  also observe  that 
the $P$-evolution pattern exhibited by the corresponding  components of 
$Q_{\rm twist-2} (y,P)$
is rather different from the nonperturbative evolution 
of PQDs $Q(y,P)$ in the models considered in our recent paper
\cite{Radyushkin:2016hsy}.  Furthermore, the comparison of the two cases
indicates that the $M^2/P^2$ target-mass corrections in   $Q_{\rm twist-2} (y,P)$ 
are much smaller than the  $\Lambda^2/P^2$  higher-twist corrections 
in our model PQDs $Q(y,P)$. 

According to Ref. \cite{Radyushkin:2016hsy}, 
  the PQDs are completely determined by the transverse momentum dependent 
distributions ${\cal F}(x, k_\perp^2)$.  Thus, our next goal is to
find the twist-2 part  ${\cal F}_{\rm twist-2}(x, k_\perp^2)$  of the total TMD.
Using the formalism of virtuality distribution functions   (VDFs) 
\cite{Radyushkin:2014vla,Radyushkin:2015gpa}
we find the   explicit form of such a TMD [it coincides with
the  results of earlier studies  \cite{Barbieri:1976rd,Ellis:1982cd}
based on a particular on-mass-shell Ansatz for the  parton-hadron blob $\chi (k,p)$]. 
The form of  ${\cal F}_{\rm twist-2}(x, k_\perp^2)$ 
 is fully specified   by the  PDF $f(x)$, and 
its  $k_\perp^2$-support  
is limited by $k_\perp^2 \leq x(1-x)M^2$.  
As a consequence, the average transverse momentum 
induced by such a TMD is rather small. In particular, for 
a toy PDF $f(x) = (1-x)^3$, it is given by $\langle k_\perp^2 \rangle = M^2/30$,
which is about $(170\,  {\rm MeV})^2$ in case of the nucleon,
that   is considerably smaller than a folklore value of $(300\,  {\rm MeV})^2$. 

Our further study shows  that the twist-2 part is not the only source of 
 kinematic target-mass corrections: they 
also come from the higher-twist
contributions.  After   incorporating 
the analysis  of target-mass dependence  for Feynman diagrams in the 
\mbox{$\alpha$-representation} and studying  equations of motion for 
the full TMD ${\cal F}(x, k_\perp^2)$, we conclude that 
${\cal F}(x, k_\perp^2)$  should depend  on $k_\perp^2$ through the combination
$k_\perp^2+x^2 M^2$, and that this is the  only  
 ``kinematically required''  target-mass effect for the full TMD. Making this modification 
 in the models used in Ref. \cite{Radyushkin:2016hsy}, we observe that
 these  $M^2/P^2$-corrections may be neglected
 well before the PQDs  closely approach the limiting PDF  form.

 The paper is organized as follows.  
In  Section  2,   we start with  recalling the definition of PQDs and their  relation to VDFs 
established in Ref. \cite{Radyushkin:2016hsy}.
Then,  using the $\alpha$-representation,
 we analyze the target-mass dependence of Feynman diagrams.
 In Section 3, 
we investigate the $M^2$-dependence of the twist-2 part of the PQD $Q (y,P)$.
Using the VDF formalism, we also find the \mbox{twist-2}  parts of the relevant VDF and TMD.
In section 4, we study the $M^2$-dependence of higher-twist
contributions and equations of motion for TMDs. 
Since the basic relations between various types of  parton distributions
 are rather insensitive to complications brought in  by spin,
 in Sections  \mbox{2 -- 4}   we refer to a simple scalar model.
 In Section  5,   we discuss modifications related to  quark spin 
 and gauge nature of gluons in quantum chromodynamics (QCD). 
 In Section 6, we discuss the  $k_\perp^2 \to k_\perp^2+x^2 M^2$
 modification of  models for soft TMDs used in Ref. \cite{Radyushkin:2016hsy}, 
and 
   present  numerical results  for  nonperturbative evolution 
 of PQDs obtained in this  modeling.   Summary of the paper and our conclusions are given in Section 7.

 \setcounter{equation}{0}   


\section{Quasi-Distributions} 

\subsection{Definition of PQDs and their  relation to VDFs} 

The   parton quasi-distributions originate from  equal-time bilocal operator 
formed from two fields  $\phi(0) \phi (z)$ separated  in space only   \cite{Ji:2013dva}, which 
corresponds  to 
$z= (0,0,0,z_3)$ [or, for brevity, \mbox{$z=z_3$}]. Then the PQDs 
  are defined  by 
 \begin{align}
  \langle p |   \phi(0) \phi (z_3)|p \rangle 
=  & 
\int_{-\infty}^{\infty}   dy \, 
 Q(y, p_3) \,  e^{i y  p_3 z_3 } \, 
 \ . 
 \label{newVDFxzQ}
\end{align}

In our paper 
\cite{Radyushkin:2016hsy} 
we have analyzed the PQDs in the context of a general
 {\it VDF representation}   \cite{Radyushkin:2014vla,Radyushkin:2015gpa} 
 \begin{align}
 \nonumber 
  \langle p |   \phi(0) \phi (z)|p \rangle & \equiv B(z,p)
=   
\int_{0}^{\infty} d \sigma \int_{-1}^1 dx\,  %
 \Phi (x,\sigma; M^2) \, \\ & \times \,  e^{-i x (pz) -i \sigma {(z^2-i \epsilon )}/{4}} \,  
 \
 \label{newVDFx}
\end{align} 
(where $M^2=p^2$)  that  basically reflects the fact that the matrix element 
$\langle p |   \phi(0) \phi (z)|p \rangle$  depends on $z$ through 
$(pz)$ and $z^2$, and may be treated as a double Fourier representation 
with respect to these variables.

The  VDF representation holds for any  $p$ and $z$, 
but it is convenient to take the frame  in which $p =\{ E, 0_\perp, p_3=P \}$. 
When $z$ has only  the minus component $z_-$,  the matrix element 
 \begin{align}
  \langle  p |\phi (0) \phi(z_-)  
| p \rangle =
   \int_{-1}^1 dx \, f(x) \, 
e^{-ixp_+ z_-} \,  \  
 \label{twist2par0}
\end{align}
is parameterized by the parton distribution function (PDF) $f(x)$ 
that depends on the fraction $x$ of the target momentum 
component $p_+$ 
carried by the parton. 
The   relation between the VDF $\Phi  (x, \sigma) $ and the collinear  twist-2  PDF $f(x)$
is formally given by 
\begin{align} 
  f  (x) = 
\int_{0}^{\infty}  \Phi (x,\sigma) \, d \sigma   \ .
\label{Phix0}
\end{align} 
The $\sigma$-integral diverges when $\Phi (x,\sigma)$ has a $\sim 1/\sigma$
hard part generating perturbative evolution of  PDFs.
Our primary concern is nonperturbative evolution,
so we will always imply  the soft part of $\Phi (x,\sigma)$
for which the $\sigma$-integral converges.

If we take $z$ having just the third component,  $z=z_3$, we have 
 \begin{align}
  \langle p |   \phi(0) \phi (z_3)|p \rangle 
=  & 
\int_{0}^{\infty} d \sigma \int_{-1}^1 dx\,  %
 \Phi (x,\sigma) \, 
 e^{i x p_3 z_3 +i \sigma z_3^2/{4}} \,  .
 \label{newVDFxz3}
\end{align} 
This gives  a relation between PQDs and VDFs,
 \begin{align}
 Q(y, P)  = & \,\int_{0}^{\infty} d   \sigma \sqrt{\frac{i \, P^2}{\pi \sigma}}
 \int_{-1}^1 dx\,  %
 \Phi (x,\sigma) \, 
  e^{- i (x -y)^2 P^2 / \sigma }
 \ . 
 \label{PQDPhi}
\end{align} 
For large $P$,
we have 
 \begin{align}
  \sqrt{\frac{i\, P^2}{\pi \sigma}}
  e^{- i (x -y)^2 P^2 / \sigma } =  \delta (x-y) + \frac{\sigma}{4 P^2} \delta'' (x-y) + \ldots
 \
 \label{Qin3}
\end{align} 
and $Q(y, P\to \infty)$ tends to the integral  (\ref{Phix0})  producing $f(y)$. 

The deviation of $Q(y, P)$  from $f(y)$ for large  $P$ 
may be described by higher-twist corrections in powers 
of $\Lambda^2/P^2$ (where $\Lambda$ is a scale like average 
primordial transverse momentum) and target mass corrections
in powers of $M^2/P^2$. 

As shown in our paper  \cite{Radyushkin:2016hsy} , PQDs are
completely determined by TMDs, so building models 
for TMDs we generate evolution patterns showing how  
$Q(y, P)$  may depend on $P$ due to the transverse-momentum effects. 

\subsection{Target mass dependence of VDFs}

\begin{figure}[t]
    \centerline{\includegraphics[width=2.5in]{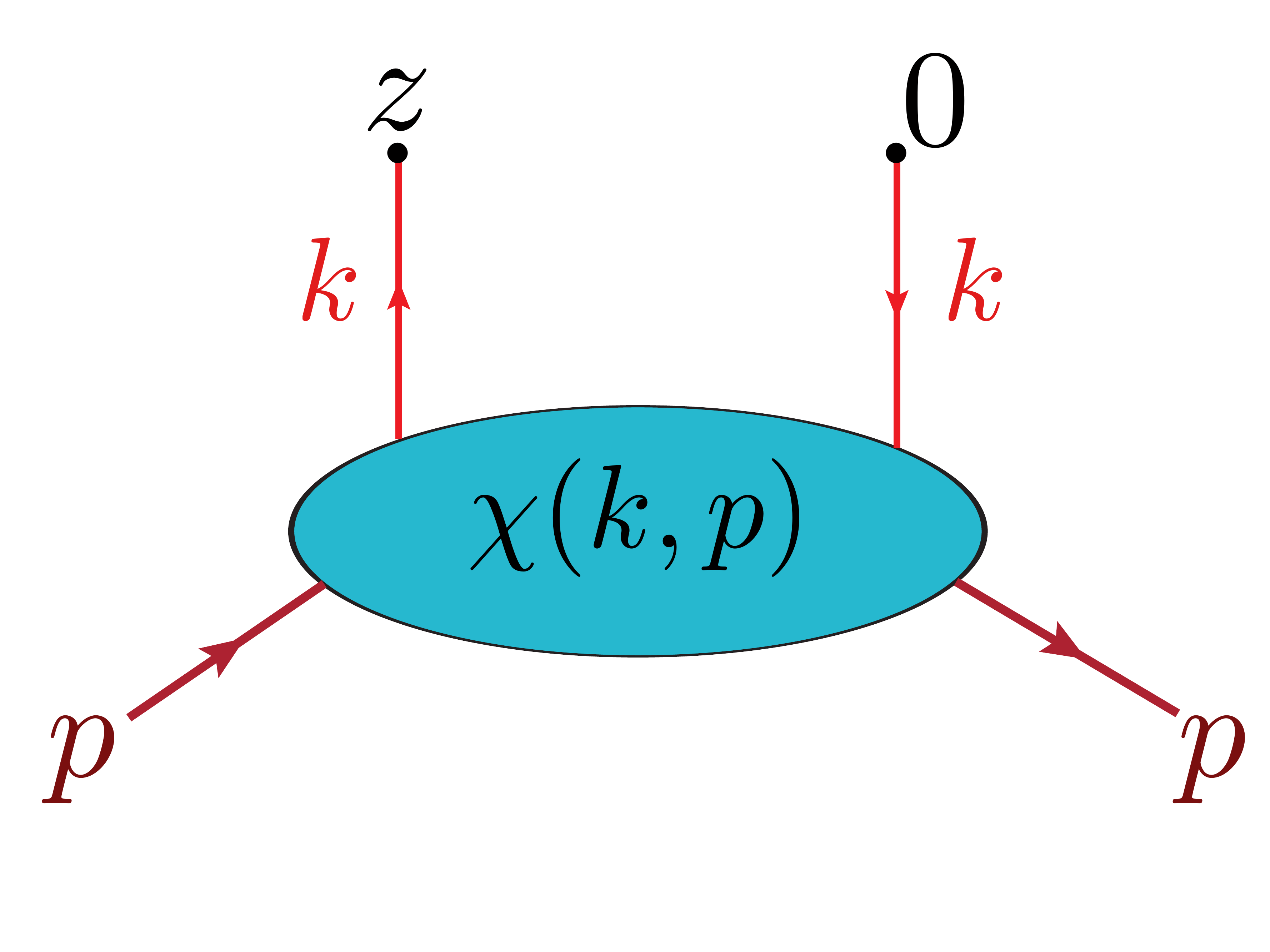}}
    \caption{Structure of parton-hadron matrix element.
    \label{chipk}}
    \end{figure}

To discuss the origin of the target-mass dependence of VDFs 
it is convenient to switch to the 
  momentum space description of the bilocal matrix element
\begin{align}
 \langle  p | \phi (0)  \phi(z)  | p \rangle = \frac{1}{\pi^2}  \int { d^4 k}  \, e^{- ikz} \,   \chi (k,p)  
 \label{twist2parz}
\end{align}
 in terms of   the function $\chi (k,p)$  (see Fig. \ref{chipk}) which  is an analog of 
the Bethe-Salpeter  amplitude \cite{Salpeter:1951sz}.

 A crucial observation is that  the contribution of 
 any  (uncut) diagram  to $\chi (k,p) $
 may be written as  
\begin{align}
  &i  \chi_{d_i}  (k,p) =  i^{l} \, \frac{P({\rm c.c.})}{(4\pi i)^{2L}}
\int_0^{\infty} \prod_{j=1}^l   d\alpha_{j} [D(\alpha)]^{-2}
\nonumber \\ & \times 
\exp \left \{ i k^2  \frac{A (\alpha)}{D(\alpha) } +i 
  \frac{ (p-k)^2 B_s (\alpha)+   (p+k)^2 B_u (\alpha) }{D(\alpha) }
 \right \} 
\nonumber \\ & \times 
\exp \left \{ 
i  p^2  \frac{ B_{p^2} (\alpha) }{D (\alpha) }
- i  \sum_{j} \alpha_{j} (m_{j}^2- i\epsilon) \right \}  
\label{alphap0}
\end{align}
(see, e.g., \cite{Nakanishi:1971graph}) , 
where    ${P({\rm c.c.})}$ is the relevant
 product of  coupling constants, 
 $L$ is the number of loops of the diagram,  and $l$
is the number  of its
 lines. The  functions
$A(\alpha), B_s(\alpha),  B_u(\alpha),  C(\alpha), D(\alpha)$ 
are   sums of products
of the non-negative $\alpha_j$-parameters.  
Using Eq. (\ref{alphap})  we get the representation 
\begin{align} 
& i \chi (k,p)  = \int_{0}^{\infty} d\lambda  \,  \int_{-1}^{1} d x \,
e^{i  \lambda[  k^2 - 2x (kp) +i\epsilon]} F(x, \lambda;p^2)  
\label{chisca} 
\end{align}
 with a  function $F(x, \lambda;p^2)$  specific 
for each diagram 
\begin{align}
&F_{d_i} (x, \lambda;p^2)    =  i^{l} \, \frac{P({\rm c.c.})}{(4\pi i)^{2L}}
\int_0^{\infty} \prod_{j=1}^l   d\alpha_{j} [D(\alpha)]^{-2}
\nonumber \\ & \times 
\delta \left ( \lambda -  \frac{A(\alpha)+ B_s(\alpha) +  B_u(\alpha)}{D(\alpha) }  \right ) 
\delta \left ( x -  \frac{B_s(\alpha) -  B_u(\alpha)}{A(\alpha)+ B_s(\alpha) +  B_u(\alpha)} \right ) 
 \nn & \times 
\exp \left \{ 
 i p^2  \frac{  B_s (\alpha)+    B_u (\alpha) +B_{p^2} (\alpha) }{D(\alpha) }
- i  \sum_{j} \alpha_{j} (m_{j}^2- i\epsilon) \right \}   \  .
\label{alphap}
\end{align}

 Transforming  Eq.  (\ref{chisca})   to the coordinate 
representation and changing $\lambda =1/\sigma$  gives
 \begin{align}
  \langle p |   \phi(0) \phi (z)|p \rangle&  
=   \int_{0}^{\infty} {d\sigma} \,
  \,   \int_{-1}^{1} d x \, 
  e^{-i x (pz) -i \sigma {(z^2-i \epsilon )}/{4}} \nn & \times 
e^{-i  x^2 M^2/\sigma} F  (x, 1/\sigma\, ; M^2)
	  \,  . 
 \
 \label{newTpqx2}
\end{align} 

Note that the quadratic dependence on $x$ in the exponential
was produced by the  $k \to z$  Fourier transformation:  originally  all terms 
in the exponential of Eq. (\ref{chisca}) 
have  linear dependence on $x$. Basically,  one gets $-x^2 M^2$
after  manipulating  $k^2-2x (kp) $ into \mbox{$ (k-xp)^2 - x^2 M^2$.}

Absorbing the factor $\exp [-ix^2  M^2/\sigma]$ into $F  (x, 1/\sigma)$
and defining  the   {\it Virtuality Distribution Function} 
 \begin{align}
\Phi  (x, \sigma; M^2) = 
\exp [-ix^2  M^2/\sigma] F  (x, 1/\sigma; M^2)
\label{Phidef}
\end{align} 
gives  the  VDF representation (\ref{newVDFx}).

Taking $z$ that has 
 $z_-$ and $z_\perp$ components only, i.e.,
 projecting on the light front $z_+=0$,  
 we define   the {\it Transverse Momentum Dependent Distribution} 
 in the  usual way 
  as a Fourier transform
with respect to remaining  coordinates   $z_-$ and $z_\perp$. 
The TMD may be written in terms of VDF as  
\begin{align}
&{\cal  F} (x, k_\perp^2) =  \frac{i }{\pi} 
\int_{0}^{\infty} \frac{d \sigma }{\sigma} \, 
 \Phi (x,\sigma; M^2) \,  \,  
 e^{- i (k_\perp^2-i \epsilon )/ \sigma} \  .  
\label{TMDsig} 
\end{align} Since $ \Phi (x,\sigma; M^2) $  {\it must} have the
$\exp [-ix^2  M^2/\sigma]$ factor, the TMD 
${\cal  F} (x, k_\perp^2)$ {\it must}  depend on $k_\perp^2$
through the \mbox{$k_\perp^2 +x^2 M^2$}  combination. 
Thus, this part of the $M^2$-dependence is  
 kinematical, and hence  predictable if we know the $k_\perp^2$
 dependence of  ${\cal  F} (x, k_\perp^2) $. 
 
In addition,  $ \Phi (x,\sigma; M^2) $, and hence also  ${\cal  F} (x, k_\perp^2) $ 
have a  ``dynamical''  or ``kinematically unpredictable''
$M^2$-dependence contained in
$F  (x, 1/\sigma; M^2)$ that comes from the last line in the 
$\alpha$-representation (\ref{alphap}).

 \setcounter{equation}{0}  

\section{Target mass dependence  of  the twist-2 part}

\subsection{PQD for twist-2 part}

Another (and well-known)  example of the kinematical
target mass  dependence  is given by the $M^2$-structure 
of the matrix elements of the twist-2 local operators. 

To get the twist-2 part of the bilocal operator 
$ \phi(0) \phi (z)$,  one should start with the Taylor expansion 
in $z$ and then change the product of derivatives 
$\partial ^{\mu_1} \ldots \partial ^{\mu_n} $
into its traceless part $\{\partial ^{\mu_1} \ldots \partial ^{\mu_n} \}$. 
In a short-hand notation $(z\partial)^n \to \{z \partial \}^n$, and 
$(zp)^n \to \{zp \}^n$, so that 
 \begin{align}
  \langle p |   \phi(0) \phi (z)|p \rangle |_{\rm twist-2} 
=  &\int_{-1}^1 dx \, f(x)   \sum_{n=0}^\infty (-ix)^n \, \frac{\{z p \} ^n}{n!} \  .
\label{tw2sum}
\end{align} 
Note that for $z=z_-$ we have $\{zp \}^n =  (zp)^n$, which  reproduces 
Eq. (\ref{twist2par0}).  To proceed  in a situation with $z^2 \neq 0$, we use the fact that 
the structure of $\{z p \}^n$  is related to  the Gegenbauer polynomials
$C_n^1 (\cosh \theta)$  equal  to  Chebyshev polynomials 
$U_n (\cosh \theta) =\sinh ((n+1) \theta)/\sinh \theta$.  As a result, 
\begin{align} 
\{zp \}^{n}  = (zp)^{n} \frac{ [1+r ]^{n+1} -[1-r ]^{n+1} }{2^{n+1}  r} 
\  ,
\label{zpnR}
\end{align}
where 
$r=\sqrt{1 -z^2 p^2 /(zp)^2} $ (see, e.g., Ref. \cite{Radyushkin:1983mj}).
Using  
\mbox{$p=(E, 0_\perp, P)$} and taking $z=z_3$,
we have  \mbox{$(zp) = -z_3 P$,}   $z^2= -z_3^2$  and $p^2=M^2$.  Thus,  we have
$$r=   \sqrt{1+ M^2/P^2  } =E /P$$
and 
\begin{align} 
\{zp \}^{n} =   (-1)^n z_3^n  \frac{ [P + E  ]^{n+1} -[P -  E ]^{n+1} }{2^{n+1}  E } 
\  .
\label{zpnR2}
\end{align} 
This gives 
 \begin{align}
 & \langle p |   \phi(0) \phi (z_3)|p \rangle |_{\rm twist-2} 
=  \int_{-1}^1 dx \, f(x)  \nonumber \\ & \times 
 \left [ \frac{E+P}{2E} e^{ ixz_3(P+E)/2} +
  \frac{E-P}{2E} e^{ ixz_3(P-E)/2} \right ] \ ,
\end{align} 
and we get the twist-2 part of PQD in the form 
 \begin{align}
 Q_{\rm twist-2}   (y, P)  = & \,   
  \frac{1}{1+  2 \Delta } \left [ f \left ( y / (1+ {\Delta }\right ) + 
f \left (- y /\Delta  \right )  \right ] 
 \ , 
 \label{newVDFzQin2}
\end{align} 
where 
$$
\Delta = \frac{E-P}{2P} = \frac{M^2}{4 P^2} + \ldots  \ \  . 
$$
This result  was originally obtained (in  somewhat different way and  notations)
in Ref. \cite{Chen:2016utp}. 
As noticed there, the 
integral over $y$ is preserved 
\begin{align}
 \int_{-\infty}^\infty dy\,  Q_{\rm twist-2} (y, P) =  \int_{-1}^1 dx\,  
f (x)  \  .
 \
 \label{Qin815}
\end{align} 
One can check that the momentum sum rule  also holds, 
\begin{align}
 \int_{-\infty}^\infty dy\, y \,  Q_{\rm twist-2} (y, P) =  \int_{-1}^1 dx\,  x \, 
f (x)  \  .
 \
 \label{Qin816}
\end{align}

Since the PQD $Q  (y, P) $ for negative $y$ may come both from the 
$y>0$ and $y<0$ parts of the PDF $f(y)$, it makes sense 
to split $f(y)$ in these two  parts and analyze 
PQDs coming from each of them separately. 
    For illustration, we take the same model as in Ref. \cite{Radyushkin:2016hsy}, 
namely, the function $f(x)= (1-x)^3\, \theta (0<x<1)$ resembling 
valence quark distributions. 

\begin{figure}[t]
    \centerline{\includegraphics[width=3.2in]{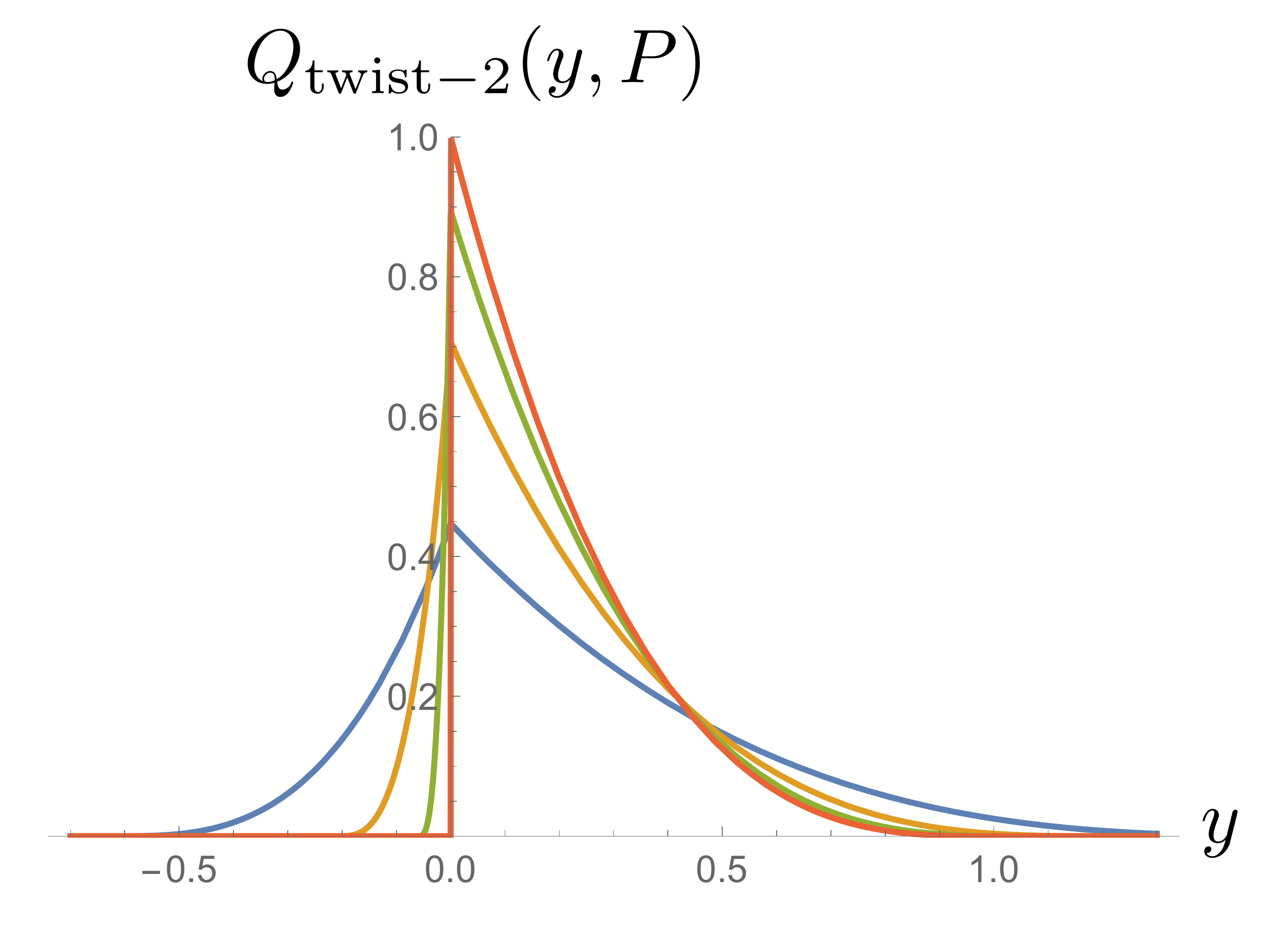}}
    \vspace{-0.4cm}
    \caption{Twist-2 part of  $Q(y,P)$   for {$P/M =0.5,\  1, \  2$}
   (from bottom to top at $y=0.1$)  compared to the  limiting PDF $f(y) = (1-y)^3 \theta (0<y<1)$.
    \label{tw2}}
    \end{figure}

According to Eq. (\ref{newVDFzQin2}), the 
  PQD for positive
$y$  is obtained  from the original $f(y)$ by stretching it  by factor $(1+\Delta)$ in the  horizontal
direction and squeezing by factor $(1+2 \Delta)$ in the vertical one (see  Fig. \ref{tw2}).
For negative $y$,  one should take $f(-y)$ and contract   it by factor $\Delta$ 
in the horizontal direction, with the same squeeze by $(1+2 \Delta)$ in the vertical one.

Thus, if  the twist-2 target  mass corrections were  the only ones here, 
it would be very easy to reconstruct such a  PDF 
from the PQD at positive $y$: one should just perform  
the $(1+\Delta)$ and $(1+2 \Delta) $ rescaling mentioned above.

For comparison, we show in Fig. \ref{Qgy} the $P$-dependence of PQD due to the nonperturbative
evolution in the  Gaussian  model of Ref. \cite{Radyushkin:2016hsy}.
Notice that the curve for $P=10\,  \Lambda$ is close in height  to the $P= M$ curve of
Fig. \ref{tw2}. 
We expect that the scale $\Lambda$ is about 300 to 500 MeV, or from 1/3 to 1/2 of the nucleon mass.
Hence, $10 \Lambda$ corresponds to about 3 -- 5 M.  One can see that 
already the $P=2M$ curve  from Fig. \ref{tw2}  
is very close to the limiting curve (in this case $\Delta =0.06$), which means that the 
target mass corrections 
in this comparison are visibly  smaller than the  higher-twist corrections governed by $\Lambda$
(despite the fact that $\Lambda$  was taken to be 2 -- 3 times smaller than $M$).

\subsection{VDF  for twist-2 part}

\begin{figure}[t]
 \centerline{\includegraphics[width=3.3in]{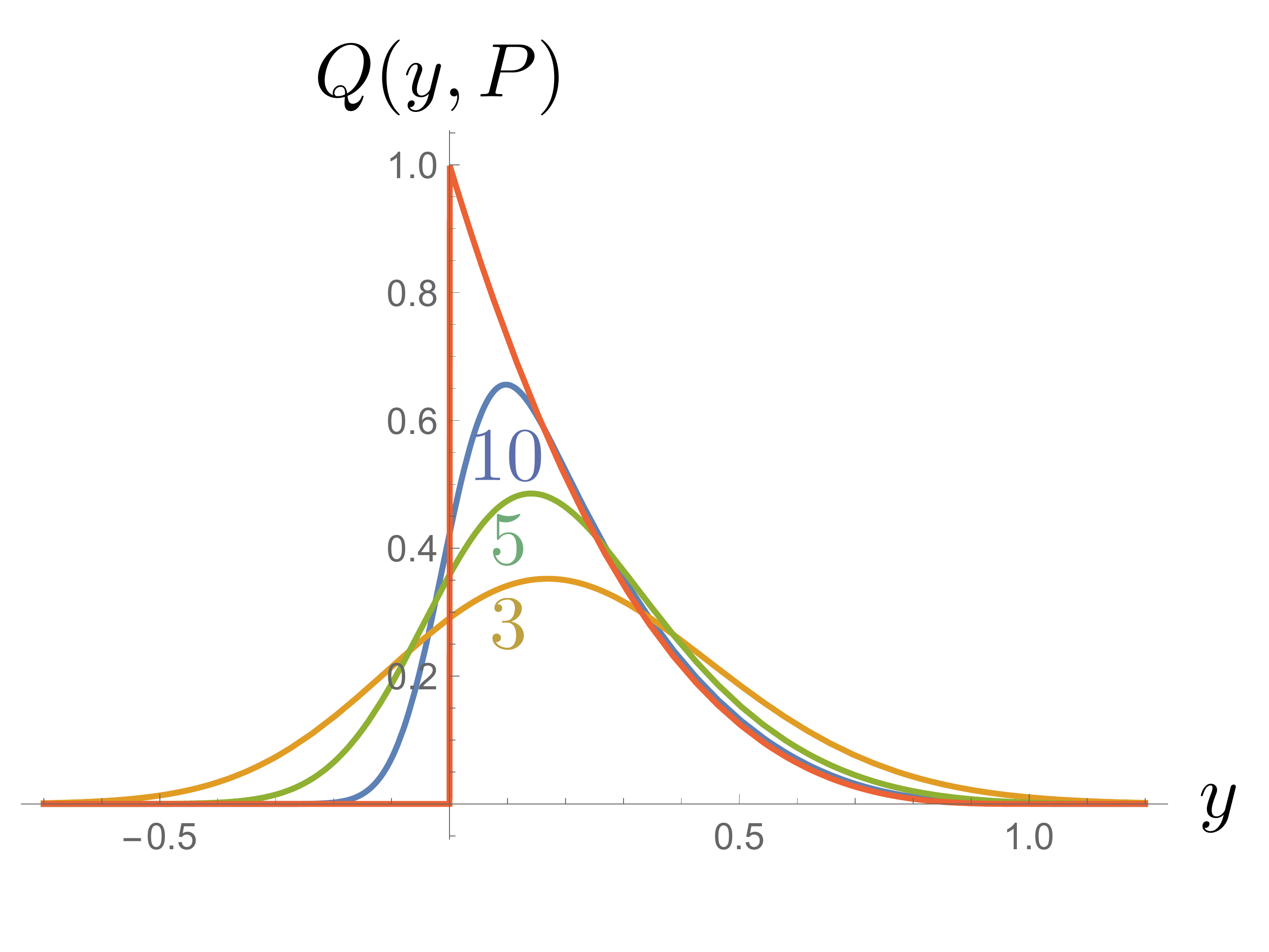}}
    \vspace{-0.6cm}
    \caption{Evolution of $Q(y,P)$  in the Gaussian  model  for {$P/\Lambda =3,5,10$}
   (from bottom to top at $y=0.1$)  compared to the  limiting PDF $f(y) = (1-y)^3 \theta (y)$.
    \label{Qgy}}
    \end{figure}

The $P$-evolution patterns in Figs. \ref{tw2} and \ref{Qgy}  are rather different.
It is interesting to find a physical reason   for  this difference.  
As shown in Ref. \cite{Radyushkin:2016hsy}, the PQDs are completely determined by 
the TMDs,
 \begin{align}
 Q(y, P)  =  & \,\int_{-\infty}^{\infty} d  k_1
   \int_{-1}^1 dx\, P\,  {\cal F} (x, k_1^2+(x -y)^2 P^2 )
  \ . 
 \label{QTMD}
\end{align} 
The Gaussian  model  mentioned above corresponds to 
a  factorized Ansatz 
  \begin{align}
{\cal F} _G (x, k_\perp^2) = \frac{f (x)}{\pi \Lambda^2}  e^{-k_\perp^2/\Lambda^2} \ . 
\label{Gaussian}
\end{align} 
So, let us find out what kind of TMD corresponds 
to the twist-2 part of the matrix element. 

The first step is to find the  VDF corresponding to the twist-2 
contribution (\ref{tw2sum}).  
To  this end, we start with      the decomposition 
   of the traceless combinations over the usual ones,
  \begin{align}
    \{pz\}^n =&(pz)^{n } \sum_{k=0}^{\lfloor n/2\rfloor} (-1)^k\frac{(n-k)!}{ k!(n-2k)!} 
    \left ( \frac{M^2 z^2 }{4(pz)^2}\right )^k    ,
   \end{align} 
   that   follows from the $\xi^k$ expansion of the Gegenbauer polynomials $C_n^1 (\xi)$.
   This gives  a double  expansion in  $(pz)$ and $z^2$  for the sum  in Eq.  (\ref{tw2sum}),  
 \begin{align}
 \sum_{n=0}^\infty  
 (-i)^n x^n  \frac{\{pz\}^n}{n!}& =  \sum_{k=0}^\infty\frac{1}{ k!}  \left (\frac{x^2M^2 z^2 }{4}\right )^k \nn & \times  \sum_{N=0}^{\infty }
 (-i)^{N} x^{N}  \frac{(pz)^N} { N!} \frac{(N+k)!}{(N+2k)!} \ .
\end{align}
 Representing 
\begin{align}
\frac{(N+k)!}{(N+2k)!} =\int_0^{1} dt_k  \ldots 
\int_0^{t_2} dt_1 t_1^{N+k}
 \end{align} 
 we get 
 \begin{align}
  \sum_{n=0}^\infty 
 (-i)^n x^n  \frac{\{pz\}^n}{n!}  & =  \sum_{k=0}^\infty\frac{1}{ k!}  \left (\frac{M^2 z^2 }{4}\right )^k 
\nn & \times \int_0^{x} du_k  \ldots 
\int_0^{u_2} du_1 u_1^{k}e^{-iu_1 (pz)}  \  . 
  \end{align} 
  At this stage, it is convenient to treat $x>0$ and $x<0$  parts of $f(x)$ separately.
  For definiteness, we take  $x>0$.  Notice now that 
   \begin{align}
   \int_0^1 dx f(x) & 
\int_0^{x} du_k  \ldots 
\int_0^{u_2} du_1 u_1^{k}e^{-iu_1 (pz)}  \nn &
=
\int_0^1 dx \,x^{k}e^{-ix (pz)} 
f_k (x)\ , 
  \end{align} 
  where the functions $f_k (x)$   are  defined by the recurrence relation
    \begin{align}
f_{k+1} (x) =   & \int_x^1 dy\,  f_k(y)
\ , 
  \end{align} 
  with $f_0 (x) = f(x)$. 
  As a result, 
   \begin{align}
 \langle p |  \phi (0)   \phi(z )|p \rangle |_{\rm twist-2}
=& 
\int_{0}^1 dx\,     e^{-ix (pz)}  \sum_{k=0}^\infty \frac{1}{ k!} \left (x\frac{M^2 z^2 }{4}\right )^k f_k(x) \  . 
 \end{align} 
Comparing with the  VDF representation (\ref{newVDFx}), we find   
   \begin{align}
\Phi_{\rm twist-2}(x, \sigma)= \sum_{k=0}^\infty \frac1{k!}  \left (-i x M^2  \right )^k \delta^{(k)} (\sigma)f_k(x) \  . 
  \label{PhiBox00}
 \end{align}

 \subsection{TMD  for twist-2 part}
 
To proceed with the formula (\ref{TMDsig} ) producing the TMD we use 
    \begin{align}
    \int_0^\infty \frac{d \sigma}{\sigma}   
     \delta^{(n)} (\sigma) \, e^{-i (k_\perp^2-i\epsilon)/\sigma} =(-i)^n  n! \delta^{(n)} (k_\perp^2) \  , 
 \end{align} 
 which results in  the $\delta^{(n)} (k_\perp^2) $ expansion 
      \begin{align}
    {\cal F} _{\rm twist-2} (x,  k_\perp^2) =&  \frac1{\pi} \, 
    \sum_{n=0}^\infty \
(-xM^2)^n   \delta^{(n)} (k_\perp^2) \, f_n (x)  \ 
\label{FckExp}
 \end{align} 
that   is equivalent to the following  expression  for the $(k_\perp^2)^n$ 
  moments of  ${\cal F}_{\rm twist-2} (x, k_\perp^2) $:
   \begin{align}
  \int_0^\infty d k_\perp^2 \, (k_\perp^2)^n  {\cal F}_{\rm twist-2} (x, k_\perp^2) = 
   \frac1{\pi} \, 
  (xM^2)^n n! f_n(x)  \ .
\label{MnFonshell0S}
 \end{align} 
 It is easy to check that the moment relation (\ref{MnFonshell0S}) is satisfied  by the function
 \begin{align}
    {\cal F} _{\rm twist-2}(x, k_\perp^2) = - 
\frac1{x\pi M^2} \,   \   
f' (x+  k_\perp^2 /xM^2)
\ .
\label{tw2f}
 \end{align} 
  In the $M=0$ limit, we have 
  \begin{align}
    {\cal F} _{\rm twist-2}(x, k_\perp^2) \Big |_{M\to 0}  =\frac1{\pi} \,  f(x) \, \delta (k_\perp^2) \ .
\label{tw2M0}
 \end{align} 
Thus, no transverse momentum is 
 generated in the case of a massless target. 
 Our illustration model  $f(x) = (1-x)^3$   gives 
  \begin{align}
    {\cal F}^{\rm mod}  _{\rm twist-2}(x, k_\perp^2) = 
\frac3{\pi (xM^2)^3} \,   \   
 (x\bar x  M^2 -  k_\perp^2 )^2 \, \theta (k_\perp^2 \leq x \bar x  M^2) 
\ ,
\label{tw2TMD}
 \end{align} 
 where $\bar x \equiv 1-x$. 
   One can  check that using the TMD (\ref{tw2TMD}) in the TMD$\to$PQD conversion 
  formula (\ref{QTMD})  one obtains the PQDs dictated by  Eq. (\ref{newVDFzQin2})
  and shown in Fig. \ref{tw2}.

The interpretation of  the twist-2 approximation  in terms of the 
  transverse momentum dependent function given by \mbox{Eq. (\ref{tw2f})} 
is known    \cite{Barbieri:1976rd,Ellis:1982cd} 
 from the early days of the $\xi$-scaling approach
\cite{Georgi:1976ve}.  
 It  was derived 
  by  imposing the  $k^2=0$ condition on the 
  hadron-parton blob $\chi (k,p)$   through the Ansatz 
  \begin{align}
\chi^A  (k,p)  =& -2 \pi 
\frac{\delta (k^2) }{M^2} f' \left ( \frac{2(pk)}{M^2}\right )   \   ,
\label{xiAn}
\end{align}
while    keeping  the  target mass finite $p^2=M^2$, 
  see, e.g.,  \mbox{Ref. \cite{Barbieri:1976rd}.}   
  In a similar context, Eq. (\ref{tw2f})  was obtained in Refs.
\cite{Efremov:2009ze,Efremov:2010mt} (see also \cite{Efremov:2015cdb}).
  
  Our VDF-based derivation 
  shows that the twist-2 TMD (\ref{tw2f})   can be obtained without  
  additional assumptions.   
  
  \subsection{Comparing TMDs} 
  
 Note that, because the support of $f(x)$ is $0\leq x \leq 1$,
 the \mbox{twist-2}   TMDs  (\ref{tw2f}) vanish  for $k_\perp^2\geq x \bar x  M^2$. 
 This should be contrasted with the usual expectation
 (incorporated  into  our TMD models in Ref. \cite{Radyushkin:2016hsy} ) 
 that TMDs  are smooth functions of $k_\perp$ with a support extending to $k_\perp^2= \infty$.

From a physical point of view, the twist-2 part   ${\cal F} _{\rm twist-2}(x, k_\perp^2) $  
describes a  situation when a free massless quark 
happens somehow to be bound within    a system with a total mass $M$. 
This  results 
in 
 a  kinematic transverse momentum
 described by a rather artificially-looking TMD of Eq.  (\ref{tw2TMD}) type. 
 Clearly, this is just a  model construction mimicking 
 a hadron by a combination of free quarks 
 with   the total  invariant mass $M$. 
 
 Comparing TMDs, 
 it  is instructive to calculate the average transverse momentum 
 \begin{align}
\| f \| \, \langle k_\perp^2  \rangle \equiv & \, \int_0^1 dx \int d^2 k_\perp\, k_\perp^2  {\cal F} (x, k_\perp^2)  
 \label{kavGen}
\end{align}
 that they induce. Here, 
\begin{align}
\| f \|  \equiv   \int_0^1 dx  f (x) \ . 
 \label{fno}
\end{align}
For $f(x)= (1-x)^3$, we have
\begin{align}
{\langle k_\perp^2  \rangle} _{\rm twist-2}  = \frac{M^2}{30} \approx  (170\, {\rm MeV})^2 \  .
 \label{kavn2}
\end{align}
For comparison, the Gaussian TMD (\ref{Gaussian}) 
gives ${\langle k_\perp^2  \rangle} _G = \Lambda^2$. 
Thus,  taking $\Lambda=M/3$   we  should  expect that $\Lambda^2/P^2$ corrections 
for PQD in the Gaussian model are  about 3 times larger 
than the $M^2/P^2$  corrections in the twist-2 part of the PQD.
This observation explains the difference between Figs. \ref{tw2} and \ref{Qgy}. 

Note that for more realistic valence PDFs $f(x)$ that are singular for
$x=0$,  the value of ${\langle k_\perp^2  \rangle} _{\rm twist-2}$ is even smaller.
In particular, for $f(x) =(1-x)^3/\sqrt{x}$   it equals to $M^2/66$, resulting 
in ${\langle k_\perp^2  \rangle} _{\rm twist-2}\approx  (116\, {\rm MeV})^2 $,
i.e. factor of 8 smaller than the expected folklore value of 0.1\, GeV$^2$. 
 
  A  rather exotic form of the twist-2 part of the TMD contradicts a  natural 
expectation 
 that TMDs  should be  smooth functions of $k_\perp$ with an unlimited support.  
 To produce  such a smooth TMD 
 (having, moreover,   a much larger ${\langle k_\perp^2  \rangle}$),
 the higher-twist terms should literally wipe out the features brought in  by  the twist-2 term.
 This is only possible if the higher-twist   terms  also 
 have the $M^2$-dependence.
    
 \setcounter{equation}{0}  
 
 \section{Higher-twist contributions} 
 
 \subsection{Twist decomposition}

 The twist-2 contribution appears as the first term in the twist  decomposition 
 of the original bilocal operator 
 \begin{align}
 \phi (0) \phi(z) = \sum_{l=0}^{\infty} 
 \left ( \frac{z^2 }{4}  \right )^l  \sum_{N=0}^{\infty} 
 \frac{N+1}{l!(N+l+1)!}  \,
 \phi(0) 
\{z{\partial} \}^{N}
 ({\partial}^2)^l 
 \phi(0)  \  , 
  \label{expandphi}
\end{align}
 (see, e.g., Ref. \cite{Radyushkin:1983mj}).
 The operators containing powers of  ${\partial}^2$ have higher twists,
and  their contribution to the light-cone expansion
 is accompanied by powers of $z^2$.  For PQDs, $z^2$ would result 
 in a $1/P^2$ suppression factor, just like for  the target-mass corrections in twist-2 contribution.

 To analyze the interplay between the twist-2 and twist-4 terms, let us take the terms bilinear in $z$, 
  \begin{align}
 \phi (0) \phi(z)|_{\rm bil}  =  & 
 \frac{1}{2}  \,
 \phi(0) 
\{z{\partial} \}^{2}
 \phi(0)  +
 \left ( \frac{z^2 }{4}  \right ) 
 \frac{1}{2}  \,
 \phi(0) 
 {\partial}^2 
 \phi(0)  
 \  .  
  \label{bil}
\end{align}
For the matrix element, this gives
  \begin{align}
2 \langle p |  \phi (0) \phi(z) |p \rangle|_{\rm bil}  =  & 
 -  \,\{zp \}^{2} \,  \int_0^1 dx\,  x^2 f^{\rm soft}(x) 
  \nn & 
  +
 \left ( \frac{z^2 }{4}  \right ) 
 \,
 \langle p |  \phi (0)  {\partial}^2 
 \phi(0)   |p \rangle
 \  . 
  \label{mebil}
\end{align}
As we discussed, $\{zp \}^{2} = \left [  (zp)^{2}- {z^2M^2 }/{4} \right ]  $ contains the $z^2 M^2$ target-mass correction term.

Since a VDF contains all  information about   the $z$-dependence of the original matrix element,
it should provide  the VDF representation for the twist-4 matrix element  $ \langle p |  \phi (0)  {\partial}^2 
 \phi(0)   |p \rangle$ as well. 
 To this end, we calculate   $\Box_z  B(z,p)$ in the VDF representation
 (\ref{newVDFx})  involving the $x>0$ part
 of the VDF,  and  get 
  \begin{align}
  \Box_z  B(z,p)  = & -    \int_{-1}^{\infty} d \sigma \int_{0}^1 dx\,  \,  e^{-i x (pz) -i \sigma {(z^2-i \epsilon )}/{4}}
 \Phi (x,\sigma)   \nonumber \\ & \times 
\left [ x^2 M^2   + x\sigma (pz) + \frac{\sigma^2}{4} z^2 + 2i \sigma \right ] 
\,  .
 \label{Baa}
\end{align} 
(We remind that $p$ in the VDF representation
(\ref{newVDFx})   is the {\it actual} hadron momentum, with
$p^2=M^2$). 
Assuming a soft $\Phi (x,\sigma)$ and taking $z=0$, we get the twist-4 matrix element
 \begin{align}
& \langle p |  \phi (0)  \partial^2  \phi( 0)|p \rangle|_{\rm soft} = 
 -  \int_{0}^{\infty} d \sigma \int_{0}^1 dx\,  \, 
 \Phi^{\rm soft} (x,\sigma)  
\left [ x^2 M^2   + 2i \sigma \right ] 
  \nonumber \\ & = - M^2 \int_0^1 dx\,  x^2 f^{\rm soft}(x) + 2
 \int_{0}^1 dx\,     \int  d^2 k_\perp \,  k_\perp^2\,
 {\cal F}^{\rm soft} (x, k_\perp^2) \ . 
 \label{d2}
\end{align} 
As one can see, it contains a term which  a) is proportional to $M^2$ 
and b) is completely specified by the twist-2 PDF $f^{\rm soft}(x)$.
This means that the kinematical target-mass correction terms  $z^2M^2$   are  contained not 
only in the twist-2 part of the original  matrix element $B(z,p)$, but also 
in its higher-twist parts. 
Most importantly, when substituted   in Eq. (\ref{mebil}), this term {\it cancels}  the
$z^2 M^2$  term coming from the twist-2 part. 
As a result, we have the expression  
  \begin{align}
\langle p |  \phi (0) \phi(z) |p \rangle &|_{\rm bil}  =  
 -  \frac12 \, (zp )^{2} \,  \int_0^1 dx\,  x^2 f^{\rm soft}(x) 
  \nn & 
  +
 \left ( \frac{z^2 }{4}  \right )  \int_{0}^1 dx\,     \int  d^2 k_\perp \,  k_\perp^2\,
 {\cal F}^{\rm soft} (x, k_\perp^2) 
 \  
  \label{mebilF}
\end{align}
free of the $M^2 z^2$ terms. 

\subsection{TMD parametrization} 

A similar  result may be easily obtained in general  case 
if one expands the $\exp [- i\sigma z^2 /4]$ factor in the VDF representation
(\ref{newVDFx})  and uses the relation 
 \begin{align}  
(-i)^l
 \int_{0}^{\infty}  {d \sigma }  \, {\sigma}^l  \, 
 \Phi (x,\sigma) \,=  \frac1{l!}
 \int  d^2 k_\perp   \,k_\perp^{2l} \,   {\cal F } (x, k_\perp ^2) \, 
  \ .
  \label{sigma_k} 
\end{align}
Then one obtains the representation of the matrix  element
 \begin{align}
  \langle p |   \phi(0) \phi (z)|p \rangle & =  \sum_{l=0}^\infty \frac1{(l!)^2} 
\left(   \frac{z^2  }{4} \right )^l 
  \int_{-1}^1 dx\,  e^{-i x (pz)} \,   \,  \nn &  \times 
  \int  d^2 k_\perp \,  %
   k_\perp^{2l}  \,   {\cal F } (x, k_\perp^2 )  
 \label{newVDFx2}
\end{align} 
in terms of the TMD $ {\cal F } (x, k_\perp^2 )$.  The sum over $l$ 
gives the Bessel function $J_0$, so we may also write
 \begin{align}
  \langle p |   \phi(0) \phi (z)|p \rangle & = 
  \int_{-1}^1 dx\,  e^{-i x (pz)}   \nn &  \times 
\pi \,   \int_{0}^{\infty} \, d k_\perp^2   %
  J_0 \left ( \sqrt{ -  k_\perp^{2} z^2} \right )   \,  {\cal F } (x, k_\perp^2 )  \  . 
 \label{newVDFx3}
\end{align} 

The {\it TMD parametrizations}  (\ref{newVDFx2}) and  (\ref{newVDFx3}) provide another form 
of the \mbox{$z^2$-expansion,}  alternative to the twist decomposition  (\ref{expandphi}). 
Its advantage is that the $(pz)$-dependence comes through the plane waves 
$e^{-ix (pz)}$ producing simple powers $(pz)^n$  rather than complicated
traceless combinations $\{pz\}^n$ containing   $z^2 M^2$ target-mass 
dependent terms   that are simply  artifacts of the twist decomposition.   The TMD representation (\ref{newVDFx3})  
is especially convenient in applications to PQDs. In particular,  it directly leads 
to the TMD$\to$PQD conversion formula (\ref{QTMD}). 

One may argue that, due to   equations of motion, like \mbox{$\partial^2 \phi = \lambda \psi \phi$} 
in a scalar $\lambda \phi^2 \psi$ theory, one may write $ \langle p |  \phi (0)  \partial^2  \phi( 0)|p \rangle$ as $
\langle p |  \phi (0)  \lambda \psi (0)   \phi( 0)|p \rangle$ or  $\Lambda^2 \langle p |  \phi (0)  \phi( 0)|p \rangle$,
with $\Lambda^2$ having no visible   $M^2$-dependence, so that there is apparently 
nothing to cancel the  $M^2$-dependence 
of $\{pz \}^2$  in Eq. (\ref{mebil}). But this is exactly the  disadvantage 
of such an  approach:  the only thing it says about  matrix elements
of $\langle p |  \phi(0)  \{z{\partial} \}^{N}  ({\partial}^2)^l  \phi(0) |p \rangle  $ type is that,
compared to  the  twist-2 case,  
they have    extra $(\Lambda^2)^N$ factors of unspecified  size and properties.

Still,  it is an interesting question of how to incorporate  equations of motions
in the VDF/TMD parametrizations of 
the  bilocal matrix element.

\subsection{VDF parametrization for off-shell quarks} 

Since   quarks in the nucleon are virtual,  the  matrix element 
$B(z,p)$  does not satisfy the free-quark equation of motion $\Box_z B(z,p)=0$. 
Keeping nonzero $z$ and integrating by parts in Eq. (\ref{Baa}),  we obtain 
\begin{align}
- \, \Box_z & B(z,p) = 
\int_{0}^{\infty} d \sigma \int_{0}^1 dx\, e^{-i x (pz) -i \sigma {(z^2-i \epsilon )}/{4}}   \nonumber \\ & \times 
\, \left (x^2 M^2 -i x \sigma \frac{\partial}{\partial x}  - i \sigma^2  \frac{\partial}{\partial \sigma }  
 -i \sigma \right ) \,  \Phi (x,\sigma) \,. 
 \label{VDFz}
\end{align} 

By equations of motion, this should be equal to  the 3-body quark-quark-gluon 
  contribution. 
For example, in a $\lambda \phi^2 \psi $ scalar model, 
 this should be equal to $\langle  p |  \phi (0) \, \lambda  \psi (z)  \phi(z)  | p \rangle 
$.
Thus, building 
the VDF  parametrization for the  matrix element  of the 3-body $ \phi  \psi \phi$ 
operator in a   situation  when $\psi$ and one of the $\phi$'s  are at the same point 
(and may be treated as one field)
we should impose the condition 
  \begin{align}
 \Phi_{\phi (\psi \phi)}  (x,\sigma) 
=
 \left (x^2 M^2 -i x \sigma \frac{\partial}{\partial x}  
 - i \sigma^2  \frac{\partial}{\partial \sigma }   -i \sigma \right) \,  \Phi (x,\sigma)  \label{eqmocon}
\end{align} 
reflecting equations of motion.
For the TMDs constructed from $\Phi$'s using  Eq.  (\ref{TMDsig}) 
(with $k_\perp^2$ substituted by $\kappa^2$ to avoid 
too clumsy notations below) this gives
  \begin{align}
  {\cal F}_{ \phi (\psi \phi)} (x, \kappa^2) =
 \left (x^2 M^2 -\kappa^2\right ) {\cal F} (x, \kappa^2) + x  \frac{\partial}{\partial x}   \int_{\kappa^2}^\infty 
d  \kappa_1^2 {\cal F} (x, \kappa_1^2) \ ,
 \label{eqmocon2}
\end{align} 
or, differentiating with respect to $\kappa^2$,
  \begin{align}
 \frac{\partial}{\partial \kappa^2} {\cal F}_{\phi (\psi \phi)} (x, \kappa^2)=
 \left [ \left (x M^2 -\frac{\kappa^2}{x} \right ) \frac{\partial}{\partial \kappa^2} 
 -  \frac{\partial}{\partial x} \right ]  x  {\cal F} (x, \kappa^2)
 \ . 
 \label{eqmocon4}
\end{align}

For the twist-2 part,  when the l.h.s. of Eq. (\ref{eqmocon4}) vanishes, 
 we have seen in Eq. (\ref{tw2f}) that 
the function $x  {\cal F} (x, \kappa^2) $  depends 
 on $x$ and $\kappa^2$ through the combination 
    \begin{align}
 \eta \equiv x +\kappa^2 /xM^2 \  .
 \end{align} 
 Noticing  that 
    \begin{align}
     \frac{\partial \eta}{\partial x} =1-\frac{\kappa^2}{x^2 M^2}
      \  \  \ , \  \ \
\frac{\partial \eta}{\partial \kappa^2} = \frac1{xM^2}
  \  , 
 \end{align} 
we can rewrite Eq. (\ref{eqmocon4})  in terms of $x$ and $\eta$ variables, 
  \begin{align}
 \frac{\partial}{\partial \kappa^2}  {\cal F}_{\phi (\psi \phi) }(x, \kappa^2) =  \left [\frac{\partial \eta/\partial x}{\partial \eta/\partial \kappa^2}  \frac{\partial}{\partial \kappa^2} 
 -  \frac{\partial}{\partial x} \right ]  x  {\cal F} (x, \kappa^2)
 \  . 
 \label{eqmocon5}
\end{align}
Now, treating  $x  {\cal F} (x, \kappa^2) $ as a function
 ${\cal G} (x, \eta)$  of $x$  and $\eta$, and introducing $G_3 (x,\eta) 
 \equiv \partial {\cal F}_{\phi (\psi \phi)}(x, \kappa^2) /\partial \kappa^2 $
 we have 
  \begin{align}
 G_3 (x,\eta) = 
 \left [\frac{\partial \eta}{\partial x}   \frac{\partial}{\partial \eta} 
 -  \frac{\partial}{\partial x} - \frac{\partial \eta}{\partial x}  
  \frac{\partial}{\partial \eta} \right ]  {\cal G} (x, \eta) \ , 
 \label{eqmocon6}
\end{align}
and finally 
  \begin{align}
 G_3 (x,\eta) =  -  \frac{\partial}{\partial x}  {\cal G} (x, \eta)
\ . 
 \label{eqmocon7}
\end{align}

If   $ G_3 (x,\eta)$ vanishes, then we conclude that $ {\cal G} (x, \eta)$
must be a function of $\eta$, in agreement with Eq. (\ref{tw2f}).   
If  ${ G}_3 (x, \eta)$ does  not vanish, 
the only restriction imposed by the  equation  of motion  is Eq. (\ref{eqmocon7}).
Thus, we may 
take   any  reasonable model for the two-body function
${\cal G} (x, \eta)$  and then just  incorporate    Eq. (\ref{eqmocon7})  [or original Eq. (\ref{eqmocon})]
as a restriction that should be satisfied by the three-body function
$G_3 (x,\eta) $, when the $qGq$  contribution is included, say, in a DIS calculation. 

Of course, choosing a model for  ${\cal G} (x, \eta)$  one should take care that
the resulting $ G_3 (x,\eta)$ is also reasonable. In other words,
if one has some  information/expectations  about the form of 
 $ G_3 (x,\eta)$, one should make an effort to find 
 a form of  ${\cal G} (x, \eta)$ that would lead to the desired (or close) form
 of   $ G_3 (x,\eta)$.

An important  lesson is that, in the context of equations of motion, it is 
natural to build    models 
of TMDs  $  {\cal F} (x, k_\perp^2) $ in the form of  functions 
of  $x$ and $k_\perp^2+x^2 M^2 $.   
 This observation is in full accord with the general conclusion 
made at the end of  the Section 2 that TMDs 
$  {\cal F} (x, k_\perp^2) $ must depend on  $k_\perp^2 $ 
through the $k_\perp^2+x^2 M^2 $ combination. 

 \setcounter{equation}{0}  
 
\section{QCD}

\subsection{Equations of motion for spinor quarks}

In spinor case, one deals with the matrix element  of a 
    \begin{align}
 B^\alpha  (z,p) \equiv \langle  p | \bar \psi (0)\gamma^\alpha  \psi(z)  | p \rangle \  
\end{align}
type. 
It may be decomposed into $p^\alpha$ and $z^\alpha$ parts:
$B^\alpha  (z,p) = 2 p^\alpha B_p (z,p) + z^\alpha B_z (z,p)$.  
These parts are not completely  independent, since there are restrictions
imposed by  equations of motion. 

Consider the handbag contribution for the virtual Compton amplitude,
whose imaginary part gives the  deep inelastic scattering (DIS) cross section. It may be written as 
    \begin{align}
 T^{\mu \nu} (q,p) = -s^{\mu\nu \alpha \beta} 
\int \frac{d^4z}{2\pi^2 } \, \frac{z_\beta}{z^4} \,\tilde B_\alpha (z,p)   e^{-i(qz)} \, \, 
 \  , 
\label{Tpqzpsi}
\end{align}
where $z_\beta/z^4$  comes from   the spinor massless propagator
 \mbox{$S^c (z) = -1/2\pi^2 \slashed z /z^4$,}  $\tilde B_\alpha (z,p)= B_\alpha (z,p)-B_\alpha (-z,p)$, and  
$s^{\mu\nu \alpha \beta} \equiv -g^{\mu\nu} g^{ \alpha \beta} + g^{\mu\alpha} g^{\nu \beta} 
+ g^{\nu\alpha} g^{\mu \beta} $. 
 
 To check the electromagnetic gauge invariance, we calculate
    \begin{align}
s^{\mu\nu \alpha \beta} \, \frac{\partial}{\partial z^\mu} \,\left (
 \frac{z_\beta}{z^4} \,\tilde B_\alpha (z,p)  \right ) =&   \frac{z_\beta}{z^4} \left [
   \tilde B^{\nu \, ,\, \beta}
-\tilde B^{\, \beta\, ,\, \nu} + g^{\nu\beta} \,\tilde B^{\, \alpha}_{\, ,\, \alpha} \right ]
 \, \,  , 
 \label{EMch}
\end{align}
 where $ \tilde B^{\nu \, ,\, \beta}  \equiv (\partial /\partial z_\beta) \tilde B^\nu (z,p)$, etc.
 
 The antisymmetric term will be eliminated 
if one takes   $B_\alpha (z,p) $   to be a derivative 
$B_\alpha (z,p) = \partial_\alpha B(z,p)$ of 
some ``generating'' scalar  function  $B(z,p)$.  After that, 
$B^{\alpha}_{\, ,\alpha}  = \Box_z  B(z,p) $, and the equation of 
motion for $B^{\alpha}_{\, ,\alpha}   = 
\langle  p | \bar \psi (0)\slashed \partial   \psi(z)  | p \rangle$  
brings us  to  a study of $\Box_z B(z,p)$,   which completely 
parallels that performed in the previous section.

As for the remaining violation of the EM gauge invariance  for the DIS handbag, 
it is proportional to $\Box_z \tilde B (z,p)$, i.e., 
 we still have it, as it is  
caused by  the virtuality of the active quarks.    In a Yukawa gluon model, 
we have $\slashed \partial   \psi(z) = i g \phi (z)  \psi(z)$, 
and this violation will be  compensated when one includes   terms coming from the 3-body 
$\bar \psi \phi \psi$  diagrams, provided that one imposes 
the restriction (\ref{eqmocon7}). 

\subsection{Equations of motion in QCD}

In QCD, one should take the operator 
\begin{align}
{\cal O}_q^\alpha  (z_2,z_1; A) \equiv \bar \psi (z_2) \,
 \gamma^\alpha \,  { \hat E} (z_2,z_1; A) \psi (z_1)  \  
\end{align}
involving the 
 gauge link   $ { \hat E} (z_2,z_1; A)$  along  the straight line connecting $z_1$ and $z_2$. 
The equation of motion,  applied to the relative coordinate $z$, takes the form 
 \begin{align}
  &  \frac{\partial}{\partial z_\alpha}  \langle p |    \bar \psi(X-z) \gamma_ \alpha  \,\psi (X+z)|p \rangle  
  \nonumber \\ &  = (ig)   \langle p |    \bar \psi(z_2) \,\gamma_\alpha  {\mathfrak A}^\alpha (z_2,z_1)   \psi (z_1)|p \rangle\,
  \ , 
\end{align} 
    where 
     \begin{align}
     &
          {\mathfrak A}^\alpha  (z_2;z_1)
        = 
     ({z_1}_\nu - {z_2}_\nu) 
    \int_0^1 dt \, G^{\nu \alpha} (t (z_1- z_2)+z_2) \  .
        \end{align} 
 As a result, we have
      \begin{align}
\partial_\alpha  B^\alpha (z,p) =     &ig \langle p | \bar \psi (0) 
     {z}_\nu \gamma_\alpha
    \int_0^1 dt \, G^{\nu \alpha} (t z) \psi (z) |p\rangle \nn & \equiv
    B_{\bar \psi G \psi} (z,p)  \  .
        \end{align} 
        Taking again $B^\alpha (z,p)= (\partial/\partial z_\alpha ) B(z,p)$  reduces the equation 
        of motion to the  equation for $\Box_z B (z,p)$ involving a scalar function $B (z,p)$,
        and   we can use all the results of Section 3, since the explicit form of $B_{\bar \psi G \psi} (z,p)$
        was not essential there.

  \setcounter{equation}{0} 
\section{Modeling target-mass dependence of   PQDs} 

\subsection{$M^2$-dependence of TMDs}

Thus, if one uses the VDF/TMD representations (\ref{newVDFx}), (\ref{newVDFx3})
for matrix elements,
there   are no kinematic $z^2M^2$-corrections 
that are artifacts of expansion  over traceless $\{pz\}^n$ combinations. 
Furthermore, the PQDs are given by the conversion formula
(\ref{QTMD}), and the target-mass dependence 
of $Q(y, P)$  may only come from that of ${\cal  F} (x, k_\perp^2)$. 

According to  the general statement made at the end of Section 2,  the  TMDs 
${\cal  F} (x, k_\perp^2)$  {\it must }  depend on $k_\perp^2$
through the \mbox{$k_\perp^2 +x^2 M^2$}  combination.  
This is a ``predictable''  or ``kinematical'' target-mass dependence.

We also noted there  that ${\cal  F} (x, k_\perp^2)$ may  have 
a ``dynamical'' $M^2$-dependence due to the $M^2$-dependence 
of the underlying function $F  (x, 1/\sigma\, ; M^2)$ of 
 Eq. (\ref{newTpqx2}).   This kind of \mbox{$M^2$-dependence}  
 cannot be   derived from kinematics, and in this sense it is ``unpredictable''.
 In principle, there is nothing special in the fact that $F  (x, 1/\sigma\, ; M^2)$
 depends on the hadron  mass,  just like there is no wonder 
 that the shape of a PDF $f(x)$ may be different if the  hadron mass   would be  different.
 
 This is to say that some part of the $M^2$-dependence 
of   $F  (x, 1/\sigma\, ; M^2)$  may be absorbed into  the form of the PDF $f(x)$,
and   would not lead to $M^2/P^2$ corrections describing the difference 
between a QPD $Q (y,P)$ and its PDF $f(y)$. 
 Still, some part of the unpredictable $M^2$-dependence 
 may lead to the $M^2/P^2$ corrections, and it is a challenge 
 to build VDF models   that would  ``realistically'' reflect that part 
 of the $M^2$-dependence.
 
  Leaving this problem for future studies,
 in what follows we will investigate   the consequences 
 of the ``mandatory'' change  $k_\perp^2 \to k_\perp^2 + x^2 M^2$
 in the TMD models that have been used for generating nonperturbative 
 evolution of PQDs  in our paper \cite{Radyushkin:2016hsy}.

\subsection{Gaussian model}

  Adding the  $M^2$-dependence into  our Gaussian model 
(\ref{Gaussian})  by the $k_\perp^2 \to k_\perp^2 + x^2 M^2 $ prescription, we get 
  \begin{align}
{\cal F} _G (x, k_\perp^2) \to  \frac{f (x)}{\pi \Lambda^2}  e^{-(k_\perp^2+x^2 M^2)/\Lambda^2}
=  \frac{\tilde f (x)}{\pi \Lambda^2}  e^{-k_\perp^2/\Lambda^2} \ , 
\label{Gaussian2}
\end{align} 
where $\tilde f (x) = f(x)\, e^{- x^2 M^2/\Lambda^2}$.  Thus,  
we have a 
 simple change in the form of the PDF, $f(x) \to \tilde f(x)$,  that would not be reflected 
by  $M^2/P^2$  terms  in the difference between $\tilde Q(x, P)$ and $\tilde f(x)$.

\subsection{Simple non-Gaussian model} 

Another VDF model proposed in Ref. \cite{Radyushkin:2016hsy}, 
\begin{align} 
\Phi_m  (x, \sigma) = \frac{f(x)}{ 2i {m}{\Lambda} K_1 (2 m/ \Lambda) } e^{i \sigma /\Lambda^2  - i  m^2 /\sigma -\epsilon \sigma }  \ , 
\label{alpharDm}
\end{align}
intends to reproduce the large-$|z|$  exponential $\sim e^{-|z| m}$ fall-off  of the  perturbative 
propagator $D^c (z.m)$ of a particle with mass $m$,  while removing its  $1/z^2$ singularity 
at small $z^2$  by a ``confinement'' factor $ e^{i \sigma /\Lambda^2}$ reflecting
the  finite size of a  hadron.  
This model 
corresponds to  the  TMD  given by 
\begin{align}
& {\cal F}_m(x, k_\perp ) = f (x)\,  \frac{K_0 \left (2 \sqrt{ k_\perp^2 +m^2 } / \Lambda \right ) }{ \pi  m \Lambda
 K_1 (2 m/ \Lambda) }  \  . 
 \label{psim}
\end{align}
Using the prescription $k_\perp^2 \to k_\perp^2 + x^2 M^2$ amounts to the change
$m^2 \to m^2 + x^2 M^2$   in this model.  To avoid a two-parameter ($\Lambda$ and $m$)  
modeling,  in our paper  \cite{Radyushkin:2016hsy}  we took $m=0$.
Let us do the same here. In the context of the $m$-model (\ref{alpharDm}), 
the resulting TMD model 
\begin{align}
& {\cal F} (x, k_\perp; M ) = f (x)\,  \frac{K_0 \left (2 \sqrt{ k_\perp^2 +x^2 M^2} / \Lambda \right ) }{ \pi  xM  \Lambda
 K_1 (2 xM / \Lambda) }  
 \label{psixM}
\end{align}
 corresponds to assuming that the parton mass
$m$ is a fraction $xM$ of the nucleon mass $M$. 
This assumption does not look   absolutely  unnatural in view of the fact 
that the VDF representation (\ref{newVDFx}) involves  the plane wave factor
$e^{-ix (pz)}$  in which $p$ is the {actual}  hadron  momentum $p$ satisfying
$p^2=M^2$.

For the quasi-distribution, the model (\ref{psixM})   gives  
 \begin{align}
 Q   (y, P;M)  = &\frac{P}{\Lambda}  \,  \int_{-1}^1 dx  \, f(x) \,  \frac{
  e^{-2 \sqrt{(x -y)^2 P^2/ \Lambda^2 +x^2 M^2/ \Lambda^2}  } }{2 K_1 (2 x M/ \Lambda)  {x M}/\Lambda}  \,
\  .
 \
 \label{Qinm}
\end{align}

\subsection{Numerical results}

Now we have two parameters, the nucleon mass $M$ and 
the transverse momentum  scale $\Lambda$, and we need to decide
what is their ratio. To this end, we calculate the 
average transverse momentum in the model of Eq. (\ref{psixM}) with $f(x)= (1-x)^3$, 
and find 
\begin{align}
\sqrt{\langle k_\perp^2  \rangle}_{\rm mod}  \approx 
\Lambda \, \left (\frac{M}{\Lambda} \right )^{0.18} \  , 
 \label{kavn}
\end{align}
with 1.5\% accuracy in the interval $1.5 <M/\Lambda <5$, i.e., for $\Lambda$ between 200
and 600 MeV.   The factor $\left ({M}/{\Lambda} \right )^{0.18}$ changes from 1.1 to 1.3
in this region. Thus, the average  transverse momentum is predominantly determined by
$\Lambda$.   Assuming  a folklore  value of 300 MeV for the average $k_\perp$,
we take $M/\Lambda=3$.

\begin{figure}[t]
    \centerline{\includegraphics[width=3in]{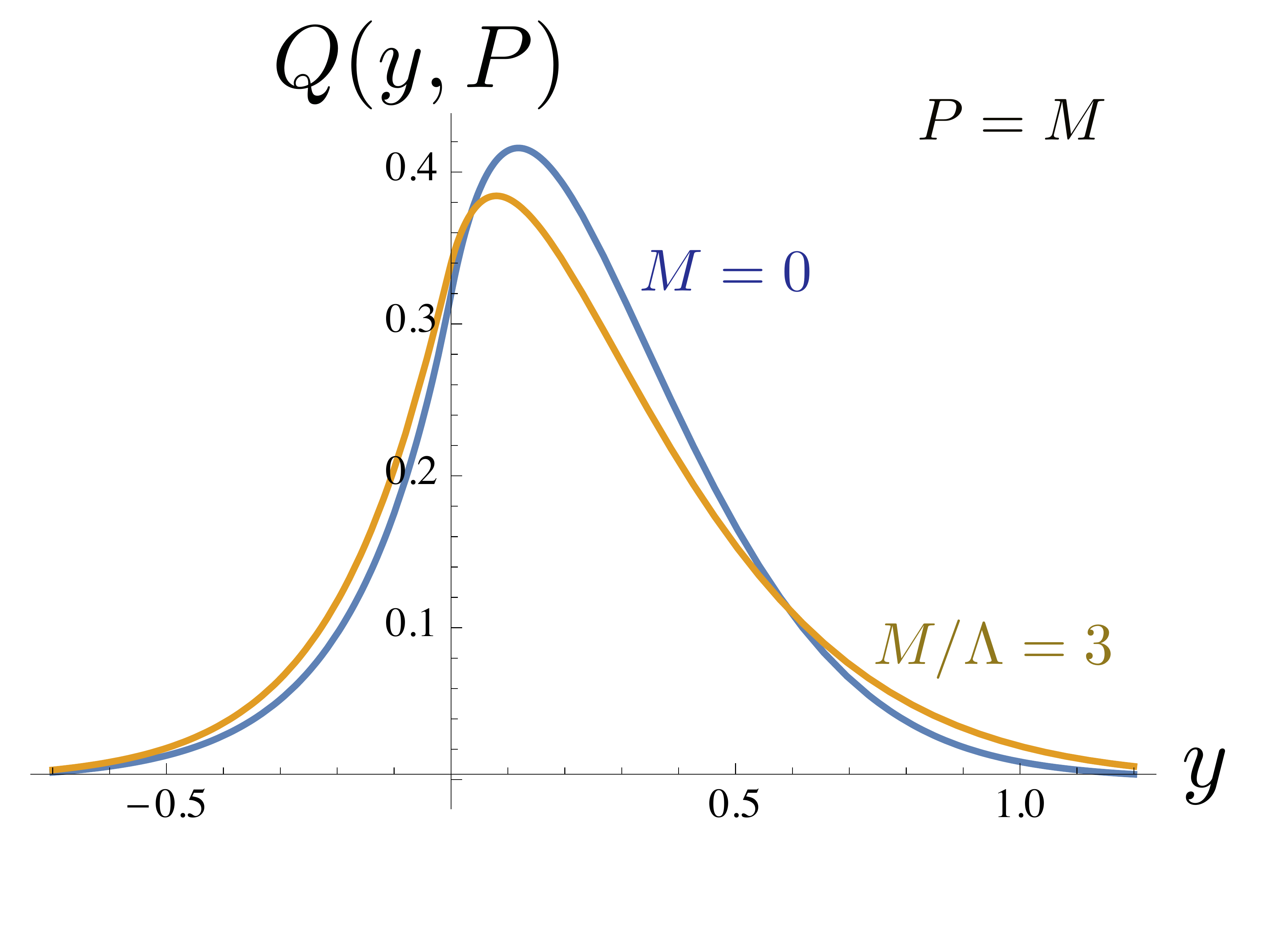}}
    \caption{Comparison of QPDs $Q(y,P;M)$ for $M=3 \Lambda$ 
    and $M=0$ at $P=M$.
    \label{fmx}}
    \end{figure}

\begin{figure}[t]
    \centerline{\includegraphics[width=3in]{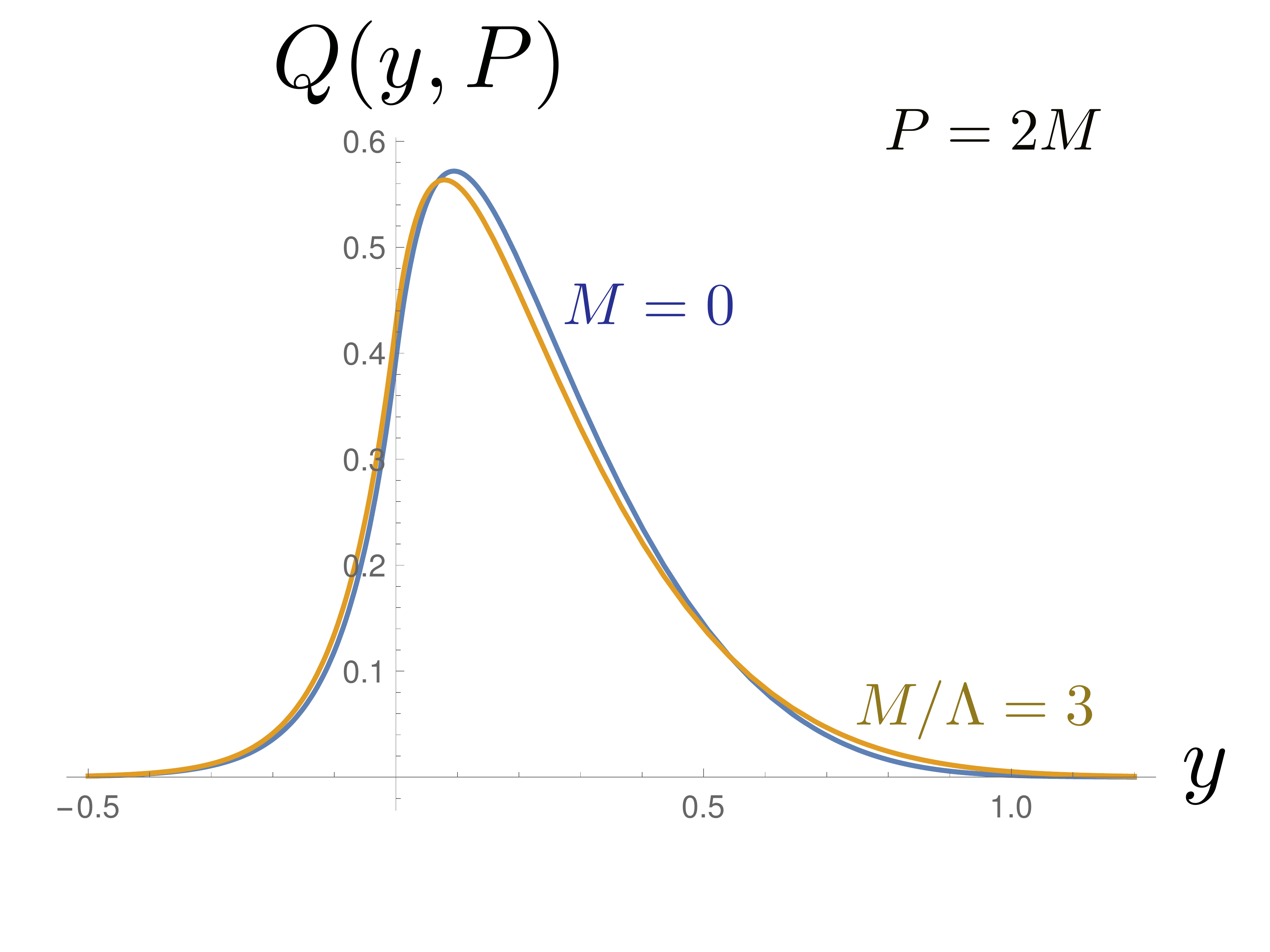}}
    \caption{Comparison of QPDs $Q(y,P;M)$ for $M=3 \Lambda$ 
    and $M=0$ at $P=2M$.
    \label{fmx2}}
    \end{figure}

To illustrate the  impact  of the  $M^2$ terms in Eq. (\ref{Qinm}) 
on the shape of 
 quasi-distributions, we  take again  
$f(y)=(1-y)^3 \theta (y)$,    and compare curves for $M/\Lambda=3$ and
$M/\Lambda=0$ at $P/\Lambda=3$ (i.e. $P/M=1$).
As one can see from Fig. \ref{fmx}, the two curves are very 
close  to each other.  At the same time, they are still very far 
from the limiting $(1-y)^3$ shape.  Increasing $P$ to $2M$,
we get the curves that practically coincide (see  Fig. \ref{fmx2}), 
still being rather far from 
the asymptotic $P/\Lambda  \to \infty$ shape.

    Thus,  in this scenario,  when one reaches  the momentum $P$ that 
     is sufficiently  large to stop  the nonperturbative 
    evolution of the  PQD $Q(y,P)$,
 there is no need to bother about   target mass corrections. 
    Given the expected accuracy of lattice gauge calculations,
    they  may be safely neglected starting with  $P\sim 2 M$. 
    
\section{Summary and conclusions}

In this paper, we have studied the target-mass dependence  of 
the parton virtuality distributions.
Our main result is that if one uses the VDF/TMD 
representations (\ref{newVDFx}), (\ref{newVDFx3})
for matrix elements,
there   are no kinematic $z^2M^2$-corrections  that are an inherent feature/artifact 
of 
 expansions   over traceless $\{pz\}^n$ combinations 
 that appear in the twist  decomposition. 
In our approach, the PQDs are given by the  TMD$\to$PQD  conversion formula
(\ref{QTMD}). In the $P\to \infty $  limit  of the latter, the PQD $Q (y,P)$ 
tends to the twist-2 PDF  $f(y)$ irrespectively of the fact that the 
VDF/TMD 
representation  does not involve the twist decomposition. 

We have established that TMDs  ${\cal F} (x, k_\perp^2; M^2 )$ must depend 
on $k_\perp^2$ through the $k_\perp^2+x^2 M^2$ combination.
Hence, the $x^2 M^2$ addition here may be considered 
as a kinematic target-mass correction. Furthermore,
TMDs may have a dynamic \mbox{$M^2$-dependence} 
that cannot be predicted from kinematical considerations.
Just like the form of the $k_\perp^2$-dependence of the TMDs,
this part of the  $M^2$-dependence can  only  be modeled in our approach.

We have studied the  effect of the $k_\perp^2 \to k_\perp^2+x^2 M^2$
modification of the TMD models used in our paper 
\cite{Radyushkin:2016hsy}, and found that the 
$M^2/P^2$ corrections become negligible 
well before the  PQD curves $Q(y,P)$ become 
close enough to the corresponding PDF $f(y)$. 
Thus, we see no need to correct the lattice gauge
calculations of PQDs for $M^2$-effects. 

A similar analysis of the target-mass effects can be made 
for the pion quasi-distribution amplitude 
studied  recently on the lattice  in Ref. \cite{Zhang:2017bzy}
and in the VDF approach in Ref. \cite{Radyushkin:2017gjd}.
Since the pion mass $m_\pi$ is much smaller than 
the nucleon mass $M$ (even when $m_\pi$ is taken in its lattice version
$m_\pi \sim$ 310 MeV \cite{Zhang:2017bzy}),
while the pion size scale $\sim 1/\Lambda$ is not very different 
from   that 
of the  nucleon, the target-mass effects in that case may be completely ignored.

A possible future extension     of our findings 
is an  
application  of the VDF/TMD approach to inclusive DIS, with the goal 
to investigate if the target-mass corrections
described there by the Nachtmann  \mbox{$\xi$}  variable 
\cite{Nachtmann:1973mr} 
are a genuine feature of the process  or just
an artifact of the  twist decomposition.

\section*{Acknowledgements}

This work is supported by Jefferson Science Associates,
 LLC under  U.S. DOE Contract \#DE-AC05-06OR23177
 and by U.S. DOE Grant \#DE-FG02-97ER41028.

\end{document}